\documentclass[sigconf]{acmart}

\AtBeginDocument{%
  \providecommand\BibTeX{{%
    \normalfont B\kern-0.5em{\scshape i\kern-0.25em b}\kern-0.8em\TeX}}}

\usepackage[T1]{fontenc}
\usepackage{booktabs}
\usepackage{graphicx}

\usepackage{xspace}
\usepackage{quoting}
\usepackage{framed}
\usepackage{multicol}

\newenvironment{packed_item}{
\begin{itemize}
  \setlength{\itemsep}{1pt}
  \setlength{\parskip}{0pt}
  \setlength{\parsep}{0pt}
}{\end{itemize}}

\copyrightyear{2020}
\acmYear{2020}
\setcopyright{acmlicensed}\acmConference[19th Workshop on Privacy in the Electronic Society, WPES'20]{}{}{November 9, 2020, Virtual Event, USA}
\acmPrice{15.00}
\acmDOI{10.1145/3411497.3420219}
\acmISBN{978-1-4503-8086-7/20/11}

\title{Proximity Tracing in an Ecosystem of Surveillance Capitalism}
\author{Paul-Olivier Dehaye}
\affiliation{PersonalData.IO}
\affiliation{MyData Geneva}
\author{Joel Reardon}
\affiliation{\institution{University of Calgary}}
\affiliation{\institution{AppCensus, Inc.}}



\begin{abstract}
Proximity tracing apps have been proposed as an aide in dealing with the
COVID-19 crisis. Some of those apps leverage attenuation of Bluetooth beacons
from mobile devices to build a record of proximate encounters between a pair of device
owners. The underlying protocols are
known to suffer from false positive and re-identification attacks.
We present evidence that
the attacker's difficulty in mounting such attacks has been overestimated. Indeed,
an attacker leveraging a moderately successful app or SDK with Bluetooth and
location access can eavesdrop and interfere with these proximity tracing
systems at no hardware cost and perform these attacks against users who do not
have this app or SDK installed. We describe concrete examples of actors who would
be in a good position to execute such attacks.
We further present a novel attack, which we call a biosurveillance attack,
which allows the attacker to monitor the exposure risk of a smartphone user who
installs their app or SDK but who does not use any contact tracing system and
may falsely believe that they have opted out of the system.

Through traffic auditing with
an instrumented testbed, we characterize precisely the behaviour of one such
SDK that we found in a handful of apps---but installed on more than one hundred
million mobile devices. Its behaviour is functionally indistinguishable from a
re-identification or biosurveillance attack and capable of executing a false
positive attack with minimal effort. We also discuss how easily an attacker could acquire a position conducive to such attacks, by
leveraging the lax logic for granting permissions to apps in the Android
framework: any app with some geolocation permission could acquire the necessary
Bluetooth permission through an upgrade, without any additional user prompt.
Finally we discuss motives for conducting such attacks.
\end{abstract}

\begin{document}

\maketitle

\section{Introduction}
Bluetooth-based proximity detection has been rapidly developed and deployed as a
means to fight the public health crisis posed by COVID-19\@. The principle is
to use mobile phones to digitally assist contagious disease contact tracing by
making a digital record whenever two phones are physically proximate. Bluetooth
is potentially suitable for this purpose because it is, in general, used for short-range radio
communication. This is used to detect and record a proximate encounter which can
be used to indicate  a risk for disease exposure for one party to an encounter
if the other party is later labelled as testing positive. Privacy is an immediate
concern. Some protocols, including the
widely deployed Google and Apple collaboration Google-Apple Exposure
Notification (GAEN)~\cite{GAENG, GAENA}, promote themselves as being safe for
users' privacy.

The technical mechanism of Bluetooth-based proximity tracing is that users
voluntarily broadcast a frequently changing \emph{pseudorandom}
identifier, known as an ephemeral id (EphId).  These EphIds are intended
to be \emph{unlinkably anonymous}, so that when a broadcaster begins to
broadcast a new one, receivers are unable to associate the new broadcast with
old broadcasts.
In typical implementations, broadcasts are \emph{linkable} to the same broadcaster with a knowledge
of a trapdoor function, such as the master key used to derive them or a central
authority who prescribes them.  This
trapdoor function is used to identify broadcasted EphIds that should be considered as
high risk for exposure, and can be shared so as to make \emph{linkable} the otherwise
unlinkable identifiers.

Broadly speaking, there are two types of contact tracing systems: centralized and decentralized.
While not the only possible implementation, centralized system involves a central authority that maintains the trapdoor
functions for all users and can further associate it with an identifier such as
a phone number to be able to notify them if they are at risk. This means
that users' broadcasted EphIds are verinymous to this authority, but unlinkably anonymous
to others. Decentralized systems involve no central authority
that knows verinyms and trapdoor functions, but instead users choose to publish
their trapdoor function so that other users may derive their EphIds and
determine if they have an exposure. That is, all users begin unlinkably anonymous,
though some may choose to become linkably anonymous so as to publish the fact
that particular EphIds should be considered as a risk for exposure.

Decentralized systems appear better for privacy. 
Unfortunately, there are two known attacks that have significant implications for the
decentralized scheme, and we introduce a novel third one here. Note that the first two, known as
\emph{false-positive} attacks and \emph{de-anonymization} attacks also exist for the centralized scheme,
though their effectiveness and ease are reduced; we reserve a discussion for
that in Section~\ref{s:discussion}. The third attack we name \emph{biosurveillance} attack, and it affects a non-user of a decentralized system whose device has been compromised. 

All these attacks work because Bluetooth beacon broadcasts are by nature public broadcasts. Anyone else
listening to broadcasted EphIds can store information about it, such as their
current location; they can also rebroadcast it elsewhere with the goal of creating false contact
events in other users that may result in false positives. 
Proponents of
decentralized systems dismiss the attacks as difficult to do and requiring
specialized hardware~\cite{DP3Twhite}, and suggest that the legal system can mitigate their
risk~\cite{DP3Tdp}.

In this work we show that this is not the case: it is easy to perform these attacks.
Instead of requiring specialized hardware and physical presence to mount the attack, we
show that it can be done simply by placing an SDK in a popular app. This can be
by creating the app, by purchasing control over a popular app, or by paying
a developer to include the SDK. Observe that this last option is standard practice in the mobile
ecosystem~\cite{loophole}. We further found and analyze the behaviour of an SDK that already exists
on \emph{hundreds of millions} of Android phones and which is uploading
\emph{full Bluetooth scans} of nearby devices including MAC addresses as well as some Bluetooth
advertising data to their central servers.

While we have no evidence of attacks against proximity tracing being conducted,
the company that produced this SDK can trivially mount false-positive,
de-anonymization, or biosurveillance attacks.  It further gives evidence that the attacks incur
little  cost because a for-profit business is monetizing this data \emph{before}
a novel health-related purpose was added to it. That is, this type of data
is already being collected in an ecosystem of surveillance capitalism
\cite{zuboff}.
This attack vector requires neither special equipment nor a physical presence in victim country.
The potential adversary requires no specific abilities, powers, or access other than
sufficient budget to write a few lines of code and get it included in a few modestly popular
apps or one superstar app. This is well within the budget of a state-level
adversary.


The contributions of this work are the following:

\begin{packed_item}
\item We present a new attack, called a biosurveillance attack.
\item We show that implementing false positive, de-anonymization and biosurveillance attacks is
trivial at scale by using commodity hardware and recruiting smartphone users as
confused deputies in the attack.
\item We show that there exist apps installed by hundreds of millions of users
that already contain an SDK that implements a substantial amount of the biosurveillance and de-anonymization attacks.
\item We discuss how false-positive and de-anonymization attacks also affect
centralized proximity tracing systems.
\end{packed_item}

The remainder of this paper is organized as follows. In
Section~\ref{s:background} we discuss background information including proximity
tracing and the Android permissions system. In
Section~\ref{s:attacks} we describe our proposed SDK-based attacks. In
Section~\ref{s:surveil} we detail network transmissions from an SDK that is
already deployed and collecting observed Bluetooth traffic. In
Section~\ref{s:discussion} we discuss our results, including possible
mitigations, similar attacks against centralized systems, and possible attackers
and motivations. In Section~\ref{s:future} we provide future work and
Section~\ref{s:conclusions} draw conclusions.

\section{Background and Related Work}

\label{s:background}
\subsection{Nymity Slider}

Goldberg's ``nymity slider''~\cite{iang} is a way of thinking about user
privacy that is relevant for this work. The nymity slider has four positions: unlinkably anonymous,
linkably anonymous, pseudonymous, and verinymous. If a user is \emph{unlinkably
anonymous}, it means that each ``message'' that they send cannot be linked
to any other message they send. If a user is \emph{linkably
anonymous}, it means that each message they send \emph{can} be linked to
other messages, but these messages do not \emph{identify} the user
themselves. If a user is \emph{pseudonymous}, then each message they
send can be linked to some assumed identity of the user's choice, but
this identity does not itself identify the user and can be dropped at any time.
If a user is \emph{verinymous}, then every message they send is linked to the
user's true name. Verinymous does not need to represent the user's actual name
but can instead by any type of persistent identifier, such as a MAC address, which users cannot easily discard.

Users may exist at different positions on the slider, and they
lose anonymity as they go from unlinkably anonymous towards verinymous. This can
happen by design, such as users intentionally revealing their identity. It can
also happen as a result of a \emph{de-anonymization attack}, wherein an attacker
uses information to move some user on the nymity slider towards verinymous
against the design of the system, \emph{i.e.}, in violation of the system's privacy
claims. For example, a de-anonymization attack could work by attaching a verinym
to some linked identifiers; it could also simply link identifiers that were not
supposed to be linked.

Users may simultaneously exist at different positions relative to different
entities. For example, a user may be verinymous to a trusted third party, as
well as to themselves, but be unlinkably anonymous to all others. Thus, a user's
nymity is relative to another entity and can change with time, but only towards
verinymous.

\subsection{Smartphone Proximity Tracing}

Contact tracing in the context of infectious diseases is the practice of
identifying who may be at risk of contagion by referencing the contacts that
an infectious person has  recently made.  It works by identifying those who may
have been exposed based on earlier contact with a diagnosed individual, and then
warning those who may be at risk.
This was typically a manual process based on personal recollection, and
typically can identify friends, family, coworkers, etc.

Recollection-based contact
tracing, however, is unable to identify strangers who may have unintentionally
come into close contact with an infected individual, \emph{e.g.}, by sharing public
transit. For this reason, many jurisdictions are attempting to use smartphones as an
electronic aid to augment their capacity to conduct contact tracing. Smartphones assist in two complementary ways:
(i) to infer a proximate encounter between two phones, and (ii) to record the
details of this encounter for later contact tracing. This is appealing in places
where a large segment of the population already carry such devices
because it reduces the deployment cost. In a sense, this paper describes how an
attacker can similarly reduce deployment costs by exploiting smartphones, this time as attack vectors.

\paragraph{Bluetooth-Based Proximity}

Bluetooth has been used to detect and infer proximity events.  Bluetooth is a commonly
used technology for short-range wireless communication for many consumer electronics,
such as headphones or game controllers. The observation is
that Bluetooth radio can be used as a ``proximity sensor'' if two phones are
able to ``hear'' each other as they broadcast. Deployments of this involve
phones periodically broadcasting a short random ``ephemeral ID'' (EphId) that serves to
identify the encounter. Any EphIds that are ``heard'' are recorded, \emph{i.e.}
by saving the EphIds to the device's storage.

This broadcasting is done with Bluetooth Low Energy (BLE) advertisements.  These
are also called beacons or advertising beacons, and already serve other purposes
in the world. For example, they are used for location sensing and can be used to
trigger location-based functionality. They can also convey information, such as
Google's Eddystone URL beacons that broadcast a hyperlink, or Apple's iBeacons,
which broadcast device identifiers. The Google-Apple Exposure Notification
(GAEN), which implements Bluetooth-based proximity for proximity tracing, uses
a new type of advertising beacon to transmit EphIds with the same existing
technology.

\paragraph{Proximity Tracing}

With a Bluetooth sensor, users' smartphones are able to record all observed
EphIds, which corresponds to specific encounters with others they have had.
Proximity tracing is then implemented by a separate mechanism that identifies
\emph{particular EphIds} that should be considered as
at-risk for infection, because that EphId was broadcasted at a time the person was contagious
\emph{and} the attenuation of Bluetooth signal strength indicates proximity. We refer to this as \emph{proximity tracing}. For example, if a user tests positive for the disease,
they can publish all their broadcasted EphIDs over their infectiosity window and
other users can determine if they are at risk.

EphIds can be implemented in different ways but existing deployments
involve a trapdoor function that is able to \emph{link} them. That is, the
EphIds broadcasted by the same user are unlinkable \emph{unless} one is aware
of a trapdoor function that links them. For example, the EphIds can be
the output of a cryptographically suitable pseudorandom number generator: with
knowledge of the random seed, the entire sequence can then be deterministically reconstructed and
therefore any two identifiers linked. The need for linkability through the trapdoor function derives in the Google/Apple implementation from bandwidth considerations.

\paragraph{Centralized and Decentralized Systems}

A prominent dimension in proximity-based contact
tracing is the characterization
of them being either centralized or decentralized. While there is diversity in the
specific implementations, the fundamental difference involves \emph{who} is aware of the
trapdoor function and whether any form of anonymity is provided.

Centralized systems involve some entity, such as a government authority, who maintains a mapping
from an individual to their trapdoor function, and informs the individual what
to broadcast and when. If an individual becomes ill, they report all their received
EphIds to the central entity. The central entity uses the trapdoor function
to de-anonymize these reported EphIds so as to notify other potentially
at-risk individuals.

Decentralized systems, in contrast, involve individuals maintaining their own
trapdoor function, thereby choosing what to broadcast and when. If they become ill,
they may choose to publish their trapdoor function through a
centralized authority. These are broadcasted to other individuals, who can then
generate all EphIds that would have been sent to determine if they were witness
to any.
Note that the design of a decentralized system prevents linking the trapdoor
function to some \emph{verinym} that identifiers the user.

A fundamental difference between these systems is information self-determinism.
In decentralized systems, individuals---through their devices---\emph{choose}
their trapdoor and \emph{choose} whether to publish it. Another difference is
the degree of nymity:
in centralized systems all individuals are verinymous to the central entity; in
decentralized systems individuals are unlinkably anonymous unless they choose to
become linkably anonymous---for example, to publish their trapdoor function if
they become ill. That is, a centralized system has a central entity that
learns a \emph{verinym} for \emph{every} encounter that an ill person has had in
recent time, even for those who are not
themselves ill, and the central authority \emph{may} choose to inform at risk users.
In contrast, in a decentralized system \emph{everyone} learns how to link
identifiers for those who report that they are ill.


We use the Alberta TraceTogether app~\cite{abtrace} as an example of a centralized system,
which we have reversed engineered.
Users register with a phone number and receive an access code. This access code
is then registered with a device identifier to identify the installation and the
user. A central server then provides users with the list of random identifiers
to be used as EphIds, along with the time interval to use them. The authority
maintains a mapping from a user's identifier (\emph{e.g.}, a phone number) to the
trapdoor function that allows it to generate and recognize the random
identifiers. Users who become sick are instructed to upload their observed
identifiers to the authority. The authority can then de-anonymize everyone who
may be at risk, and notify them, e.g., by text message.

We use the SwissCovid app based on the Google-Apple Exposure Notification as an
example of a decentralized system. Here every user picks their
own trapdoor function to derive random identifiers. If they become sick, they
can cooperate with a public health authority who then publishes
their trapdoor function. All other users then learn the trapdoor and recompute
all EphIds that are now considered to correspond to possible encounters with a contagious
person. They do not reveal to anyone whether they have in fact witnessed this
encounter, but instead check offline---\emph{i.e.}, only by looking at their own
stored data---whether any EphIds associated with a
risk for contagion exist among their collection of encounters.

\paragraph{Privacy Risks}

The privacy intrusions for centralized systems are clear. First, all users are
verinymous to the central authority. Second, ill users upload all their
encounters, which verinymously identifies both parties to the encounter to the
central authority. This revelation occurs even if one of the parties is not
actually ill, or, in some systems, even if the encounter was considered ``at
risk''
\cite{StopCovid}. Thus, the decision to reveal the fact whether a user was
proximate to someone else is not under the user's control once they have
broadcasted their
EphIds. Additionally, users lack control over the data: it is the authority's
decision as to whether to alert users, instead of users being able to assess
their own risks based on, for example, a heightened concern by being
more vulnerable or near those who are, as well as  lifestyle factors such as
always wearing a mask when out of doors.

As such, the decentralized system is superficially a better solution for
privacy. Users are empowered to assess their own risk based on information they
collect. Only they learn about encounters they have with others. No entity knows
their trapdoor function except the user unless they intentionally share it with
others. There is no mapping from actual personal information to the trapdoor or
broadcasted beacons aside from the user. In this work, however, we show that the
privacy implications can be worse in some decentralized systems, where users who
publish their trapdoor function can be made verinymous and have their
geolocation tracked by an adversary with a modest budget.

\subsection{Smartphone Permissions}

The Android platform
uses a permission system to protect access to sensitive
resources based on the security principle of \emph{least privilege}.
Apps must request relevant permissions in order to be able
to perform specific functionality. Permissions are requested statically in a
``manifest'' file and can thus be audited when the app is installed.

An important distinction among Android permissions are those classified as
\emph{dangerous}. These include access to sensors such as the camera and GPS, access to
user data such as call logs, and access to actions such as sending text messages \cite{permissions}.
These are permissions that Android believes are dangerous if misused by apps and
users are given stronger controls over them. For example, on Android 6.0 and
later, users are \emph{asked} the first time these permissions are used, and
importantly have the ability to \emph{deny} access to the app. Conversely, so-called \emph{normal} permissions get granted upon installation of the app, or silently added if the app upgrades after the developer has expanded their manifest file. 

Crucially, the permissions that an app has access to are also granted to any
other code that runs as part of that app. This includes the prolific ads and
analytics networks that are often included for monetization of apps. These can
report back user data to servers and offer other features to developers. The use
of such third-party code is common---it is easier and generally better to make
use of existing reusable code than to rewrite everything \emph{ad hoc}.
Nevertheless, it means that some third-party code can find itself embedded in an
app that has requested the right set of permissions to perform its desired
behaviour.

For Bluetooth, there are two relevant
permissions: \textsc{bluetooth} and \textsc{bluetooth\_admin}. The former allows
the use of Bluetooth devices, while the latter permits scanning for and connecting
to new Bluetooth devices, as well as receiving and broadcasting BLE advertisings.
As such, the latter is the permission of interest for proximity tracing apps and
attacks against them. Examples of well-known Android apps benefiting from \textsc{bluetooth} and \textsc{bluetooth\_admin} permissions would include Spotify or Uber. Despite the significance of having administrative power over Bluetooth on a
device, Android considers both Bluetooth permissions ``normal''.

Bluetooth has a further toggle in the system status bar. Users can disable the
Bluetooth radio, as well as other radios and features on the device. For
scanning to work, the Bluetooth radio must be on. Note that any use of
Bluetooth, such as for headphones or car audio, requires that Bluetooth is
enabled and thus passive scanning is possible for other apps. It remains enabled
when not in use unless users diligently disable it when finish using a
Bluetooth device.

Since Android 6.0, apps are required to hold a location permission
in order to scan for nearby Bluetooth devices and nearby WiFi
routers. This is because the serial numbers for these devices can act as a surrogate for
location~\cite{wifileaks}. As such, collecting them is now considered to imply the collecting of location as well.  The location permission is considered a ``dangerous'' permission.
Users are prompted before allowing it the first time and can revoke it through a
sequence of settings options. If an update to an app adds location access, then
users are alerted to this fact before it is used.

In summary, any Android app can acquire the permissions necessary for conducting the attacks we describe here without \emph{any mention} of Bluetooth in a user prompt. At best, the user will be prompted to add some location permission if the app does not already have that one, but would manage to invisibly slip in necessary Bluetooth permissions at the same time. 

\section{Proposed Attack}
\label{s:attacks}

There are three attacks we consider: false-positive, de-anonymization and
biosurveillance attacks. All these attacks are performed by an adversary who is capable of running code
on millions of field-deployed smartphones. For example, the attack could be
performed by an SDK that an app maker includes for monetization purposes, which
is a standard practice. The key observation is that the
attack can be mounted \emph{without} specialized equipment and \emph{without} a
legal presence in the country being attacked. The attacker does not need to own
the phones themselves, but can just leverage the existing ecosystem of mobile
app monetization, where app makers are paid to include ads and analytics SDKs
that run arbitrary code on users devices.

We call this SDK the \emph{attack SDK} and assume that it is provided by the
attacker and included in a popular smartphone app or a bevy of less popular
ones. The attack SDK collects and
transmits Bluetooth data and reports all received data back to the attacker. It
also allows the device to broadcast arbitrary Bluetooth low energy advertisements.
The attack SDK results in end users who install an app containing this SDK to
act as \emph{confused deputies}: they
unknowingly undermine the integrity or anonymity of the contract-tracing
system by contributing access to their hardware to the attacker.

Note that there is no technical obstacle to creating such an SDK\@: all of the
functionality exists in the Android API\@. 
In the next section we show the
existence of an SDK---already deployed on hundreds of millions of devices---that
collects and reports back Bluetooth low energy advertising data that are
``heard''. We do this not because we believe that this SDK is mounting these
attacks,
but rather to \emph{provide evidence} that these attacks cannot be
optimistically dismissed as hypothetical or unrealistic; they are well
within the budget of many types of adversaries. For the rest of this section, we
assume that such an SDK exists and describe how to use it to mount false-positive,
       de-anonymization, and biosurveillance attacks.

\subsection{False-Positive Attack}

A false-positive attack works by compromising the integrity of the system by
creating false contact events among users. The presumed goal is to create false
positives in risk notifications and so the attacker's goal is met if the fake
contact events they generated later correspond to positive diagnoses, since this will trigger a notification. The implementation of the
attack is to rebroadcast EphIds heard in one location somewhere else, in effect,
a relay attack~\cite{DP3Twhite}.
For example, the attacker may harvest beacons from a site likely to contain COVID-19
cases, such as a COVID-19 testing facility or a hospital, and relay them
anywhere that the attacker wishes to cause a fake outbreak, such as at a factory
for an economically vital industry, or during advanced voting to force election
officials to self-quarantine and discourage citizens from voting.

In this case, the attack SDK is configured to harvest all observed Bluetooth
beacons, including the EphIds. These are then uploaded to a central server as
soon as they are available. The attack SDK is likely to get the
user's location: it is either running on a version of Android since 2015
that requires location permission to scan for Bluetooth, or it is not and can
use surrogates to GPS-based location such as cell-phone tower triangulation or
router-MAC-address-based location.
Note that these surrogates work even when GPS is disabled or unavailable, and in
the context of our attacks are of sufficient granularity to identify actionable
location information, such as ``at a hospital''.

Based on geolocation, the attacker then decides which beacons should be relayed
by having other smartphone users running the attack SDK begin broadcasting them.
These users are given the beacons to broadcast and they simply start advertising.
This can be done by having the attacker SDK contact a control server with their
location, and depending on the attacker's goal, be given some EphIds to begin
broadcasting along with a time to broadcast them.

Note that EphIds have a short range of time where they can be relayed. This is
by design of the system to both prevent tracking of users and limit the
effectiveness of a relay attack. The Alberta TraceTogether app uses a period of
15 minutes before switching EphIds. The SwissCovid app rotates EphIds every 10 minutes, but the GAEN framework accepts them as valid within a two-hour window.
The length of the window does not change the feasibility of the relay attack,
even if it is only a minute. The attacker does not need to be physically present
at either the collection site or the relay site.
Collected EphIds can be quickly telecommunicated from the mobile phones that first
received them to a central server and onward to the devices that will
rebroadcast them using the Internet. Shortening the window of time simply means
that this collection and communication occurs more often.

\subsection{De-anonymization Attack}

A de-anonymization attack works by having an attacker gain information that
is not meant to be learned by them. In our case, it refers to
users ``losing'' positions on the nymity slider for EphIds relative to an adversary.
EphIds are intended to be unlinkably anonymous, but can become linkably
anonymous if users publish their trapdoor. In GAEN, this occurs by design when a
user tests positive for COVID-19 so as to inform all other users of the EphIds
that correspond to encounters with a risk for exposure.
A de-anonymization attack is anything an adversary can do to push the
nymity slider towards verinymous, against the user's interest.

Our proposed attack involves SDKs linking broadcasted EphIds data
to other available information, such as MAC address, GPS location, and
persistent identifiers. For example, the SDK can centrally store all received
EphIds along with the precise location where they were received. Note that these
geotagged EphIds are collected by \emph{other} devices, i.e., not the one
emitting it.

If a user chooses to publish their trapdoor, they then allow any entity to link
their emitted EphIds. Any of these EphIds that have been collected and geotagged by other
nearby devices are then not only linked but further geotagged, resulting in a
location history for a user who did not have the attack SDK installed on their
device. This can include information about their routine such as where they
sleep, work, and relax, and thus reveals a great deal about users. This occurs
even if the victim user does not use any apps that collect location or even turn
on their GPS.

Additionally, the EphIds are not broadcast in isolation but instead alongside
the device's Bluetooth MAC address.
The key problem is to associate private data, such as a number that later is
used to signal a health status, with
something that never meant to be secured, such as a MAC address.
Indeed, Google and Apple recognize this: they now  randomize (see Bluetooth specification in \cite{GAENA, GAENG}) the
broadcasted Bluetooth MAC address in tandem with the changing EphId, so that
neither one can be used to bridge changes in the other. They also prevent
programmatic access to the Bluetooth MAC address~\cite{programmaticmac}: apps, and thus embedded
SDKs, cannot determine the user's MAC address. 
Despite that, the current MAC address is still broadcasted, and so privacy
relies on the privacy of a value being unknown to the entity that is the one
actually \emph{broadcasting} it to everyone else. We describe some attacks
related to this in the discussion section.

\subsection{Biosurveillance Attack}

This attack is novel, and in fact quite simple. It can be conducted
independently of a de-anonymization attack, and concentrates on inferring the
health status of the owner of a particular device. The victim need not
use any contact tracing app and even have a GAEN-aware mobile device. This means
that the user may believe that they have entirely \emph{opted out}
of participating in proximity tracing. Despite that, an attack SDK running on
such a device can simply behave like the passive ``listening'' component of the
contact-tracing system.

The attack SDK collects the observed EphIds that are broadcasted nearby---the
very same ones that  collected by a legitimate contact tracing app. They
can use the health authority's public information to determine which of these
correspond to at-risk encounters. This allows the attacker to perform their own
risk calculation about the user without the user being aware. Effectively, the
presence of ambient EphIds creates a new ``health'' sensor available to mobile
devices in the same way that ambient GPS signals creates a ``location'' sensor.


For privacy reasons, the GAEN system introduces limitations on the data it
collects and makes available.  Beacons that can readily be deemed to be too
distant will be discarded right away. Apps using the framework will only be able
to do coarse computations on beacon characteristics, and only for those beacons
that are declared infected. Users will only see the end results.

On the other hand, the attack SDK is not bound by such considerations. It can
observe distant and fleeting beacons as well as those never deemed infected. It can
further collect the time and place for these encounters.

This provides the SDK with a better view of the ambient traffic and
the prevalence of the GAEN system than the GAEN system itself. This also allows
for population-level epidemiological research, such as outbreak detection.

\subsection{Features of the Attacks}

All these attacks can be done by an attacker with \emph{no relation}, legal or otherwise,
with the country in which the attack is mounted. Neither the attack SDK nor the
app that includes it need to have any connection with the affected country for
the attack to work. It is
only the end users who unknowingly undermine the system who are physically
present in the affected country.

This is not an exotic attack that requires a high-level of sophistication or
domain-specific knowledge. The implementation effort of scanning and
broadcasting Bluetooth signals is greatly reduced with APIs designed exactly to
facilitate this kind of development.
Furthermore, for the false positive and de-anonymization attacks, there is no
need for the victim \emph{themselves} to be the one who is also running the
attack SDK---it simply needs to have a baseline presence among other mobile
phone users.


\section{Existing Surveillance of Bluetooth}

\label{s:surveil}


In this section we give evidence for the ease that these attacks could be
performed by attackers with modest means.
Neither the relay attack nor the de-anonymization attack are novel to this work;
they are inherent to the design of GAEN\@. Despite that, we feel both their
feasibility as well as our novel attack is not fully appreciated.
For example, the Swiss National Cyber
Security Centre, discussing the GAEN-based \emph{SwissCovid} app, stated that:
``There is no real safeguard with the current design
against this [relay and de-anonymization] attack vector, however there are only few
operators of such systems and they are under the Swiss
jurisdiction which gives at least some protection on the legal
level.''~\cite{NCSC}  This suggests a belief that the attack needs either
leveraging control over already deployed specialized equipment and a physical
presence to mount the attack, such as a small ground team that collects data
from one location and rebroadcasts it elsewhere. The design documents of DP-3T
\cite{DP3Twhite} also refer to the threat of relay attacks in the
context of high-energy broadcasting and high-gain antennas.

The problem with basing security on an economic argument is that attacks can
become trivial if the economics change or the cost of all possible
implementations of the attack are not considered. Our SDK-based attack is an example
of this: it performs the same attacks without requiring expensive specialized
equipment. It is mounted entirely
with commodity smartphones already  deployed; instead of high-gain antennas and
high-energy antennas, the attacker simply uses \emph{many} smartphones on the
field to receive and broadcast data.
We show the feasibility of this attack by detailing the behaviour of an
analytics SDK that is already deployed and which harvests all Bluetooth
broadcast data it observes from other users. This SDK, which exists as part of
the world of surveillance capitalism, can easily perform all three attacks.
We detail the SDK's behaviour by dynamically analysing two apps that contain the
SDK and recording the network traffic that it generates while executing.

We first describe our experimental method to test apps and collect data. We then
describe a detailed analysis of two apps that contains the same SDK that harvests 
available Bluetooth data. We then list other apps in which we found this SDK
and, by doing an intersection of listed third parties on the apps privacy policies, we
attribute it to the company X-Mode.

\subsection{Experimental Methods}
We perform our  dynamic analysis on a Pixel 3a (sargo) mobile phone. The phone
is running an instrumented version of Android Pie, which records network traffic
for later analysis, including traffic secured by TLS. Our instrumentation
attributes network traffic to the specific app that is responsible for its
transmission. It also logs access to permission-protected resources, such as
performing Bluetooth scans. As such, we are able to observe the
real-world behaviour of an app as it executes.

Our instrumented operating system further injects spurious Bluetooth scan
results when an app attempts to scan for nearby devices.
We use conspicuous palindromic MAC addresses in our injected
results and search for them being transmitted. We format the Bluetooth
advertising data to match standard beacon formats and ensure that the raw bytes
are valid for the format. We inject an
iBeacon (MAC address \texttt{AB:B1:E6:6E:1B:BA}), an AltBeacon
(MAC address \texttt{AB:B1:E7:7E:1B:BA}), an Eddystone URL (MAC address
\texttt{AB:B1:E8:8E:1B:BA}), and a GAEN beacon (MAC address
\texttt{AB:B1:E9:9E:1B:BA}).

We start the app and accept its request for permissions, and then leave
the app running with occasional random UI interactions through the Android
``monkey'' utility.
We then obtain the network traffic and process it with a suite of decoders to
remove standard encodings such as gzip and base64. We remove network traffic
from other apps, such as system ones, and consider only that traffic being sent
by the app under investigation.

\subsection{App Collection}

We used the following method to collect apps with Bluetooth permissions from the
Google Play Store. First, we search the play store using a dictionary of 4842
English adjectives. We collected the names of apps that appear as the top
results when searching these. This gave us a list of 34952 unique app package names.

We then scraped the permissions requested by each of these apps. We
were able to get this information for 28270 of them. We then filtered the list to
include only apps that requested both the \textsc{bluetooth\_admin} permission
and a location permission, i.e., either \textsc{access\_fine\_location} or
\textsc{access\_coarse\_location}.  This left us with 1358 apps that are
permissioned to perform this attack.

These 1358 represent an astounding 4.80\% of random Google Play
Store apps sampled by our method, nearly one in twenty.
Note that the newest
version of Android, which is only deployed on a small number of phones, requires
strictly \textsc{access\_fine\_location}, not \textsc{access\_coarse\_location}.
This has little impact on the prevalence of apps able to do this: of the 1358
apps, 1215 apps required the fine location permission. Recall that
\textsc{bluetooth\_admin} is not considered ``dangerous'', meaning that apps can
update to include it without alerting the user. Through our methods, we found 7835 apps with a location
permission---more than one in four (27.7\%).

We then ran each of the apps holding \textsc{bluetooth\_admin} and a location
permission on our dynamic analysis testbed based on prior
published methods~\cite{coppa}.  We examined apps to see if any transmitted
our spurious Bluetooth MAC addresses. We found a handful of apps doing exactly
this with all of the data going to one of two domains. We reversed engineered
the SDK to confirm our findings and trace the Bluetooth scanning to network
transmissions. We now give example transmissions for two  apps containing this SDK\@.

\subsection{Case Studies on Two Apps}

In this section we discuss the findings for two apps, FunDevs LLC's Video MP3
Converter (\textsf{com.fundevs.app.mediaconverter}) and PixelProse SARL's Bubble
level (\textsf{net.androgames.level}). The former app is notable as it
has more than one hundred million installations and more than 600,000
reviews with an average of 4.4 out of 5. The latter app is less popular (only
more than 10 million installations) but receives different configuration options
and also behaves differently, which is noteworthy in itself.
On June 21st, 2020, we downloaded and installed the app
\textsf{com.fundevs.app.mediaconverter} from the Google Play Store to our instrumented Pixel 3 mobile
phone and ran it with our dynamic analysis testbed. We collected all its network
traffic while we ran the app and then analysed it afterwards. We confirmed our
findings by repeating this on July 22nd, 2020.

The first contact is to
\texttt{bin5y4muil.execute-api.us-east-1.} \texttt{amazonaws.com}
(port 443) where the app performs a GET request for \texttt{/prod/sdk-settings} and
provides an API key as an HTTP header (\texttt{x-api-key}). It
returns a JSON object storing a configuration. This includes a number of
parameters for Bluetooth scanning:
\begin{packed_item}
    \item \texttt{"baseUrlDomain":"api.myendpoint.io"}
    \item \texttt{"beaconsEnabled":false}
    \item \texttt{"bleScanMaxPerHour":2}
    \item \texttt{"btScanMaxPerHour":2}
\end{packed_item}

This initial configuration retrieval also occurs for other apps by other developers
that contain X-Mode's SDK\@. The actual configurations do change, however, and
it may depend on the API key that requests it. Observe the
\texttt{"beaconsEnabled":false} flag: for the Bubble level app this value is set
to \texttt{true} and the resulting behaviour of the SDK changes. This shows how
the attack SDK could evade detection by selectively engaging in attack behaviours
only when necessary.

There are two domains that receive transmissions of sensor data:
\texttt{smartechmetrics.com} and
\texttt{myendpoint.io}.
Figures~\ref{smartechmetrics}--\ref{myend-last} show examples of these
transmissions. Both domains receive the results of a
scan of nearby WiFi routers and Bluetooth devices as well as precise GPS
coordinates. The \texttt{myendpoint.io} domain only transmits precise GPS when
\texttt{beaconsEnabled} is set to false; when it is set to true, it also sends
nearby MAC addresses as well as Bluetooth LE advertising
data as well.

Figure~\ref{smartechmetrics} presents a transmission to 
\texttt{smartechmetrics.com}.
We have redacted identifying information, and observe that some of
the transmissions of WiFi routers are devices that are
nearby to our testbed but not actually devices owned by the authors. We see that
the MAC addresses for all injected Bluetooth traffic are collected and
transmitted, along with two pieces of consumer Bluetooth electronics in the same
room as the testbed.

\begin{figure}[tbh!]
{\footnotesize
\begin{verbatim}
{"multi_part":{},
 "obs":[
  {"gaid":"<AAID>",
   "ids":["google_aid^<AAID>"],
   "lat":<LATITUDE>,
   "lon":<LONGITUDE>,
   "metadata":["device:AOSP on BullHead",
               "os_version_int:25",
               "sdk_version:1.9.2-bcn",
               "app:Video MP3 Converter"],
   "observed":[{"mac":"AB:B1:E6:6E:1B:BA","name":"null",
                "rssi":-12,"tech":"ble"},
               {"mac":"AB:B1:E8:8E:1B:BA","name":"null",
                "rssi":-12,"tech":"ble"},
               {"mac":"AB:B1:E7:7E:1B:BA","name":"null",
                "rssi":-12,"tech":"ble"},
               {"mac":"AB:B1:E9:9E:1B:BA","name":"null",
                "rssi":-12,"tech":"ble"},
               {"mac":"<NEARBY MAC>","name":"null"
                "rssi":-50,"tech":"ble"},
               {"mac":"<NEARBY MAC>","name":"<NEARBY NAME>",
                "rssi":-43,"tech":"bluetooth"},
               {"mac":"<NEARBY ROUTER MAC>","name":"<NEARBY SSID>",
                "rssi":-50,"tech":"wifi"},
             <... 7 more observations ...>],
   "timepoint":"1592782636",
   "token":"z1+y/FCqZ2ZD8QNldpJasF/te5KBhHqXT0YlT5L/eOw="}]
}
\end{verbatim}
}
\caption{Transmission payload from Video MP3 Converter to
        \texttt{api.smartechmetrics.com:443}. Advertising ID, precise
            geolocation, and nearby  MAC addresses and device names have been
            redacted. Whitespace has been added for clarity. Observed that the
            first MAC addresses in the \texttt{observed} array correspond
            to our spurious traffic.\label{smartechmetrics}}
\end{figure}

\begin{figure}[tbh]
\footnotesize
\begin{verbatim}
[
  {
    "beaconType": "EDDYSTONE_URL",
    "isCharging": false,
    "loc_at": 1592412157818,
    "mac": "AB:B1:E8:8E:1B:BA",
    "name": "",
    "rssi": -12,
    "scan_record": {},
    "tech": "ble",
    "time": 1594004410113
  },
  {
    "isCharging": false,
    "loc_at": 1592412157818,
    "mac": "AB:B1:E7:7E:1B:BA",
    "name": "",
    "rssi": -12,
    "scan_record": {},
    "tech": "ble",
    "time": 1594004410112
  },
  {
    "beaconType": "IBEACON",
    "isCharging": false,
    "loc_at": 1592412157818,
    "mac": "AB:B1:E6:6E:1B:BA",
    "name": "",
    "rssi": -12,
    "scan_record": {},
    "tech": "ble",
    "time": 1594004410112
  },
  {
    "isCharging": false,
    "loc_at": 1592412157818,
    "mac": "XX:XX:XX:XX:XX:XX",
    "name": "XXXXXXXXXXXXXXXX",
    "rssi": -53,
    "scan_record": {},
    "tech": "bluetooth",
    "time": 1592412158800
  },
  {
    "isCharging": false,
    "loc_at": 1592412157818,
    "mac": "XX:XX:XX:XX:XX:XX",
    "name": "XXXXXXXXXXXXXXXX",
    "rssi": -54,
    "tech": "wifi",
    "time": 1592412158401
  },
      < truncated >
]
\end{verbatim}
\caption{Data sent by the app \textsf{net.androgames.level} to the domain
	\textsf{api.myendpoint.io}. We added whitespace for clarity and
	    redacted identifiers.}
\label{myend-scan}
\end{figure}

\begin{figure}[tbh]
\footnotesize
\begin{verbatim}
{
 "all_beacon_data": [
  {
   "accuracy": 20.934XXXXXXXXXXX,
   "ad_id": "<AAID>",
   "altitude": <ALTITUDE>,
   "beacons": [
    {
     "distance": 0.97085,
     "layout_name": "",
     "mac_address": "AB:B1:E6:6E:1B:BA",
     "major": "53479",
     "minor": "42571",
     "mumm": "AB:B1:E6:6E:1B:BA_01022022-fa0f-0100-
              00ac-dd1c6502da1c_53479_42571",
     "rssi": -12,
     "uuid": "01022022-fa0f-0100-00ac-dd1c6502da1c"
    }
   ],
   "bearing": 0,
   "latitude": <LATITUDE>,
   "longitude": <LONGITUDE>,
   "model": "AOSP on sargo",
   "os_version": "9",
   "platform": "android",
   "sdk_version": "1.9.2-bcn",
   "speed": 0,
   "time": 1592412157818,
  "vert_acc": 2
  }
 ]
}
\end{verbatim}
\caption{Data sent by \textsf{net.androgames.level} to
	\textsf{api.myendpoint.io}. We redacted sensor data. We added
	    whitespace for clarity and broke the \texttt{mumm} value over two
	    lines. Observe that this includes the
	    transmission of the advertising data from a Bluetooth
	    beacon.} 
\label{myend-last}
\end{figure}

Figure~\ref{myend-scan} shows an example WiFi and Bluetooth scan being
sent to \texttt{api.myendpoint.io} by the app
\texttt{net.androgames.level}. We see that in addition to actual wireless
devices, all our injected Bluetooth devices are included in the transmission. The corresponding
transmission from \texttt{com.fundevs.app.mediaconverter} only sent location,
which appears to be because the command and control server gave the instruction
that \texttt{beaconsEnabled} is set to \texttt{false} while it was set to
\texttt{true} for apps that sent this information.


Figure~\ref{myend-last} shows another transmission sent to
\texttt{api.myendpoint.io} by \texttt{net.androgames.level}. This
includes not only precise geolocation and a result from a Bluetooth scan but it actually
includes the \emph{advertising data} of the Bluetooth scan itself.  The list of beacons in the
JSON transmission has one entry corresponding to the injected iBeacon. The key
\texttt{mumm} we believe stands for \emph{mac uuid major minor}, because it is
in fact an underscore-separated string of those four fields. We observe that the
uuid, major, and minor values are exactly those that we configured to be sent as the
advertised data.

From this case study of one SDK, we understand that it is
\emph{already the case today} that Bluetooth advertising data is being read,
processed, and sent to servers on the Internet by \emph{millions of
users} while they go about their day. Note that Android permits this scanning to
occur in the background, so users do not need to use the apps in question in
order for this to be collected and uploaded.
There is no technical limitation that prevents the
full collection of the advertised data: a few simple lines of code could make
them also upload the GAEN EphIds. These can be then sent out to other devices
and rebroadcasted, or accumulated to enable de-anonymization and biosurveillance attacks.

\subsection{Prevalence of SDK}

\begin{table*}[tbh]
\begin{tabular}{lllll}
\toprule
\textbf{App} &\textbf{Package} &\textbf{Installs} &\textbf{Reviews}
&\textbf{Third Parties} \\
\midrule
Video MP3 Converter     &com.fundevs.app.mediaconverter &100M+  &600K+  &Admob,
      Smaato, Mobfox, Tutela, \\
& & & & and X-Mode\\
Fleet Battle - Sea Battle       &de.smuttlewerk.fleetbattle     &10M+   &130K+ &X-Mode only\\
Bubble Level    &net.androgames.level   &10M+   &200K+  &nearly one hundred including X-Mode\\
SPEEDCHECK Internet Speed Test   &org.speedspot.speedanalytics   &10M+   &484K+ &nearly forty including X-Mode \\
Speedcheck      &org.speedspot.speedspot        &1M+    &62K+   &Opensignal,
      Sens360, Tutela, X-Mode,\\
& & & & HUQ, and Ogury\\
Compass &fr.avianey.compass     &1M+    &35K+   &nearly one hundred including X-Mode\\
Just a Compass (Free \& No Ads)         &net.androgames.compass &1M+    &21K+ &nearly one hundred including X-Mode \\
Portable ORG Keyboard   &com.audiosdroid.portableorg    &500K+  &1K+ &Opensignal and X-Mode \\
The Sun Ephemeris       &fr.avianey.ephemeris   &50K+   &1K+    &nearly one hundred including X-Mode        \\
Altimeter PRO   &fr.avianey.altimeter   &50K+   &1K+    &nearly one hundred including X-Mode\\
\bottomrule
\end{tabular}
\caption{List of apps that we found that contain the SDK that collects Bluetooth signals,
	along with number of installations, number of comments, and third
	    parties listed in the privacy policy.\label{t:intersection}}
\end{table*}

Using our testing methods, we have found this SDK in 10 apps available on the
Google Play Store. Table~\ref{t:intersection} presents the list of apps that we
found, along with the number of installs, number of reviews, and a list of third
parties. The MP3 converter app has more than 100 million installs; the next
three have more than 10 million. They are  a battleship game, an app to measure if a
physical surface is level, and an  Internet-speed testing app.

To determine who is responsible for this SDK, we studied the privacy policies of
all these apps. In particular, we manually looked for third parties or trusted parties
with whom information is shared. We found only one common entity among them and,
it was present for all of them: X-Mode. One of the apps simply hyperlinked to X-Mode's privacy policy as their entire
statement about ``Collection and Sharing of User Information''~\cite{smuttle}.

X-Mode, formerly Drunk-Mode~\cite{drunk}, is a data aggregator and manager. Their privacy policy \cite{xmodepolicy} states that they focus on precise location data for marketers, advertisers and researchers for financial and market research, for traffic and city planning, for educational purposes, but also, since May 17th 2020, for disease prevention, security, anti-crime and law
enforcement. They collect precise geolocation, duration of
time spent in places, as well as BLE sensors and beacons, IoT signals, data sent
over NFC, and persistent identifiers of users.

To be clear, we have seen no evidence that X-Mode's SDK is performing any of the false positive, de-anonymization or biosurveillance attacks, and likewise for any of the apps listed in
Table~\ref{t:intersection}. Arguably, neither their collection of user data nor
broadcasted Bluetooth data violate any laws as the behaviours are clearly
documented on their privacy policy and users have no expectation of privacy for
messages they broadcast over public radio.

Our focus on this particular SDK is to show the
depth and scale of the existing state of surveillance capitalism in which proximity tracing is being
added. In particular, that there already exists a field-deployed Bluetooth
signals scanner and aggregator, which centralizes data from
hundreds of millions of mobile devices. The owners of these hundreds of millions of
devices that do the actual work may be unaware that they are doing this
collection on behalf of X-Mode if they do not carefully read privacy policies.
This collection is presumably profitable, as X-Mode
is a for-profit company that sells this information. The consequence is that
deploying an attack SDK may even pay for itself by also monetizing the data that is
collected, significantly reducing the cost to perform the attacks described earlier.

\section{Discussion}

\label{s:discussion}
\subsection{Coverage Requirement}
The epidemiological utility of the GAEN framework roughly scales with the square $\alpha^2$ of the prevalence rate $\alpha$ of the app within the population. Since $\alpha$ is a proportion, so below 1, it is crucial these apps get significant adoption to be of any broad utility for epidemiological purposes. 
One might naturally anticipate a similarly prohibitive prevalence requirement for the attacking SDK. This is
actually not the case. Indeed, a GAEN app is constrained by design to require
both relevant devices to have the app installed, and will not do anything as
long as neither of the two relevant users is infected. By opposition, an
attacker does not have that constraint. They can eavesdrop and interfere with
any communication taking place between two other devices (and even create new
connections in the case of replay attacks). We have not precisely quantified the
requirements for the attacks described earlier, but previous research for other
real-life networks has shown that intrusions on a network originating in just one node can cause
population-level privacy loss quickly~\cite{YAM}, when a node has the capacity to observe traffic within its neighbourhood.

\subsection{Operating System Controls}

Mobile phones can prevent access to EphIds by filtering out messages meant for
proximity tracing from the rest of the Bluetooth scan results and prevent
broadcasting over BLE of messages that could be interpreted as  an EphId for any
particular proximity tracing system.
This limits the
attackers ability to perform this attack because fewer devices can be enlisted
to perform the attack. That is, the data is still being broadcasted publicly,
but there are fewer devices that the adversary can use to collect the data.

Bluetooth permissions could be restricted. The permission could be considered
dangerous and users given control with run-time prompts and the ability to
restrict Bluetooth for third party apps while still getting to use Bluetooth
headphones. This also prevents apps from updating to silently add administrative
control over Bluetooth.

A further control is to disallow scanning for Bluetooth signals for apps running
in the background. The behaviours we observed from the SDK occur even if you do
not use the app, because the Android allows apps to start silently in the
background and perform Bluetooth scans, as well as communicate to their servers,
without any user engagement with the app.  This means that apps that
have been installed and stopped being used can still exhibit this
behaviour. The devices that the attacker can use decreases precipitously if
users are required to be actually running the app that contains the attack SDK,
instead of just having once ever installed it.

All these controls should be implemented; unfortunately none will
address the issue. This is because
support for security fixes for mobile phones are shorter than the useful
lifespan of a device. Older phones can be refurbished and repurposed, and given
to people who may not be able to afford a state-of-the-art phone every three years.
Privacy updates are not given the same importance as security updates because
they are typically not patched but rather come with newer versions of the
operating system.

According to Android's distribution dashboard~\cite{distdash}, data regarding
distribution of Android devices can be found using the Android App making tool
Android Studio~\cite{androidstudio}.  When creating an app, one can specify
a targeted version to determine how many devices will support that. Using this
tool we found that 11.2\% of devices are still using Android 6.0, a version
released first in 2015 and last updated in 2017. Devices using this version do
not receive security updates and it is officially unsupported. A further 7.4\%
of devices use versions of Android even older than 6.0. These numbers are
similar to another source, StatCounter~\cite{stat}, which puts them it at 8.4\%
for version 6.0 and 6.4\% for earlier.

As long as a baseline of phones in an area run  versions of
Android with privacy vulnerabilities, an SDK can take advantage of that fact to
perform these attacks. The attacker further needs only one device in the right place to
do the broadcasting.
Given the great diversity in the Android ecosystem, including a variety of smartphone
manufacturers with customized operating systems, the ability to entirely disable this
functionality is unlikely. This is not necessarily Google's fault for
decentralizing smartphone usage instead of Apple's model of centralizing it, but
it does mean that millions of field-deployed Bluetooth sensors are easily at
command for an adversary and are out of reach of security update mechanisms.

This illustrates the difficulty in adding a security or privacy purpose to
a feature on a commodity device that was never meant to have it, versus building
systems with security by design. If broadcasting
over Bluetooth and accessing broadcasts from Bluetooth were considered dangerous
from the original implementation, there would not be countless legacy devices
for which it was accessible. Similarly, legacy devices may process observed BLE
advertisements in irresponsible ways without realizing that it now is associated
with sensitive health data.

\subsection{MAC Address Echo Attacks}

Following on this theme is the idea that Bluetooth MAC are now considered
too sensitive to allow apps to access. If the attacker records EphIds, it can
also record the current randomized  Bluetooth MAC addresses that is broadcasted
alongside. Such an adversary needs only to associate \emph{one} of these MAC
addresses to a persistent identifier belonging to the user, or to know that it
came from a particular phone. There is a certain irony to the fact that a value
ought to remain unknown to the person who is the one intentionally broadcasting it to
everyone else.

This is vulnerable to an \emph{echo attack}, where an SDK
simply repeats back everything it hears over Bluetooth, allowing the sender to
learn its identity.
The initial broadcast of MAC address and EphId is done by the proximity
tracing app, it is received by a nearby phone and the attack SDK then echos it
back using a different Bluetooth service UUID to separate it from traffic
filtered out for proximity tracing apps only, and then received by the same
attack SDK running on the original phone.

This echo attack allows the SDK to learn the user's Bluetooth MAC
address and EphIds despite operating system controls that allow access to this
information. This data can then be sent to a central service along with
the user's advertising ID or other persistent identifiers. This gives them a
mapping from verinym to MAC address from the user, and a mapping from MAC
address to EphId can be made by anyone listening to broadcasts. Such an echo
system may even occur entirely benignly.

A more aggressive attack would broadcast a persistent
identifier instead, which would then be \emph{automatically} annotated with the
Bluetooth MAC address---the same MAC address being used for EphIds. The same SDK running on
another device can upload this data to the attacker without echoing. This gives the attacker the
same pairs of mappings to de-anonymize users.

This de-anonymization attack is \emph{substantially} harder than the one presented in
Section~\ref{s:attacks} because it requires two
phones in close proximity running the same attack SDK, and only users running
the attack SDK can become victims. It would require nearly 71\%
($\frac{1}{\sqrt{2}}$) of the GAEN-using 
population to have the attack SDK such that half of encounters could be mutually
de-anonoymized. Privacy-conscious individuals are likely to not install
closed-source apps that use surveillance capitalism for monetization and thus
greatly reducing their risk.
Note, however, that the attack works even if these two phones encounter each
other \emph{only once} during the period of relevance of the trapdoor function.
If a user broadcasts their trapdoor function, the attacker needs only
be able to map \emph{one} EphId to a verinym through a MAC address in order to
make verinymous all broadcasted EphIds.

\subsection{Attacks against Centralized Systems}

While we discussed the attacks in the context of decentralized systems, it is
important to note that many of the same risks apply to centralized systems as well.

\paragraph{Relay Attack}

First, the
relay attack can still be executed as described. Without any additional
scrutiny of reported EphIds, false positives will affect the integrity of
the system by misleading the central authority, who then issues erroneous
exposure alerts.

The central authority, however, has additional tools at their
disposal because of their high-level view of all reported infections. For example,
collecting identifiers at one site and broadcasting them widely across
a geographic area may reveal the attack through manual proximity tracing efforts,
\emph{i.e.} that different individuals were in different places at the time yet still
observed the same beacon. This
increases the cost on operating the system by requiring extra analysis and
scrutiny while having little cost for the attacker, and can  erode trust in the
intent of the authorities as individual users may not perceive the necessity
of such monitoring.  Moreover, the attacker can mitigate risk of exposure by
only relaying the signals at one specific target instead of more broadly.

Even if the central server is able to correctly identify all EphIds involved in
relay attacks, this opens up a \emph{false negative} attack for centralized
systems. Instead of creating fake outbreaks, the attacker relays all EphIds with
the goal of getting them ignored by the central authority. This prevents users
from actually receiving exposure notifications and degrades the utility
of the proximity tracing app.

A central authority could also encode geographic information in EphIds. For
example, different lists of EphIds could be given to the user with instructions
on which set to use depending on some rough location, such as which cell phone
tower is nearest. For example, a user may select an EphId from their list such
that the hash of it and a nearby cell tower's identifier have, say, four zeros at the
start.  Receivers will reject EphIds that do not pass this check, limiting the
locations where it can be relayed. This further adds geolocation information to the data collected by
the central authority. 

\paragraph{De-anonymization Attack}
The de-anonymization attack, however, is less effective for centralized systems.
The attacker can still
implement an echo attack to link broadcasted EphIds to a persistent
identifier. They cannot, however, learn the specific health status of that
individual because that information is not published. The attacker is also
unable to ``query'' the database because the system works by notifying users
only when they appear on \emph{other} users' submissions.  Some attacks are
possible. For example, the attacker could broadcast an EphId only once to a
single victim; if that victim uploads their data then the attacker may get
notified that they are at risk and know that it could only have come from the
victim.

Another risk is an \emph{insider threat}. If an attacker had access to the
centralized service mapping EphIds to verinyms, then this information may get
abused. Such a system also presents a single point of failure that can have
enormously devastating consequences if it is accessed or published without
authorization and erode public trust in similar future initiatives.


If the central authority abuses their power by using data about contacts by
people for any other purposes than the intended proximity tracing, then a new attack is available: the attacker can
try to frame a user by creating fake contact events. For example, the attacker
collects EphIds from their target, and broadcasts it elsewhere with the
goal of having other users report this fake encounter to the central authority
when uploading all their encounter data. 
This may have consequences depending on how this data is abused by the
government. For example, the encounter may represent the user in
violation of a bail condition,  in violation of a lockdown requirement, or
simply  present at an event that a politically-restrictive government has
banned.

\paragraph{Biosurveillance Attacks} By design, centralized systems do not distribute any of the health status data necessary for the risk calculation. Therefore biosurveillance attacks are not possible as described here, unless the SDK learns of infected persons through other means. 

\subsection{App Stores and App Permissions}

Apple and Google both have tremendous powers acting as gate keepers for their
respective mobile app marketplaces. Google is frequently criticized,
particularly in academic work, but it is worth noting that Apple's lack of
criticism reflects its status as a closed society. Open societies, like
democracies, invite criticism and provide transparency. Android's open source
operating system and computer control over mobile devices permits automated
analysis of app at scale that results in them being criticized precisely because
it is comparatively easy to do so. Anyone with an Android device can audit its
security and perform large-scale analysis.

Apple, in contrast, offer a limited number of ``positions'' in a research device program subject
to a legal contract with Apple, an application review, and
only for those who are membership account holders in the Apple Developer Program
and have ``a proven track record of success in finding security issues on Apple
platforms, or other modern operating systems and platforms''~\cite{applesec}. The
device remains Apple's property and participants are required to report findings
about Apple products to Apple. They stress that not all qualified applicants
will be accepted due to the decision to limit manufacturing of devices that facilitate scrutiny.

Both can make many improvement on how apps are recommended to users.
Apps with invasive permission requirements
should not be as highly recommended as similar apps without them. Basic static
analysis can detect if permission use is likely necessary only for third-party libraries.
Users should be able to search for apps by specifying permissions they are
unwilling to grant. Finding open-source apps that do not serve ads, collect
user data, or even use the Internet, should be made much easier. Internet usage
should also be considered ``dangerous'', so users cannot be misled that an app
that uses all their sensors is safe because---for now---it does not report it.
Users should also be afforded mechanisms to deny Internet access to apps that
have no legitimate need for it, even if that means only open-source apps remain
as interfaces for basic sensors such as flashlight and compass.

As noted, nearly \emph{five percent} of a random sample of Google Play Store
apps had both access to a user's location and \emph{administrative} control over
Bluetooth. We leave a deeper scrutiny of this as future work, but to give some perspective we would still like to give now a random sample of ten apps from our list of apps pulled from the Play Store (sample obtained by running the \texttt{shuf} command followed by
\texttt{head -n 10} on the whole set). In the sample are a carwash reward
loyalty app; a cafe app for a particular cafe in the Detroit community, an app
for an arts and craft store, a piano metronome (whose privacy policy makes no
mention of Bluetooth and for which there is no associated hardware), an app for
an awards ceremony, and a few
games.  The single app with obvious need to scan for Bluetooth was the ``Happn''
app,  which uses Bluetooth-based \emph{proximity tracing} to recommend people to
date with whom you keep crossing paths.
Control over the
administration of Bluetooth devices should be reserved for a small number of
apps that need that privilege; not compass and bubble level apps that happen to harvest
this information without utility to their stated purpose. Indeed, it is exactly
this type of abuse that a least-privilege-based permission system is meant to
thwart.

Google should consider both Bluetooth permissions \emph{dangerous} so that apps
cannot add them without user warning, and that users can granularity disable
access to Bluetooth on a per-app basis. The global toggle for Bluetooth is insufficient as
some users may have a Bluetooth peripheral that they require using.
Apps with associated hardware should have the
ability to engage in user-aided pairing through operating system intents
rather than requiring administrative control and thereby granting that privilege
to any embedded SDK that comes with it.

\subsection{Potential Adversaries and Motives}

Our adversarial model has two weak assumptions: (i) they can write code or pay someone
to write it, and (ii) they created a popular app, can purchase control of it,
or can incentivize an app maker to include their SDK for monetization purposes.
There are no technical limitations or powers that the adversary needs to mount
this attack; they do not need to compromise any systems; having a budget is
sufficient. The software they need to write is straightforward, with all the
APIs and frameworks to implement the attack
readily available.
The false positive attack is the only attack described here that requires active
interference with the system and has the potential to be detected and classified
as malware; the other two attacks can be implemented offline by analysing data
that is already being collected.

The existence of a vulnerability, however, does not imply an attack: we require
a motivated attacker.
To better assess the risks and thus the threat of this attack, we consider
all the adversary types from Van Oorschot's categorical schema~\cite{van2020basic}.
It divides adversaries into foreign intelligence, politically-motivated
adversaries and cyber terrorists, industrial espionage agents, organized
criminals, lesser criminals, malicious insiders, and non-malicious employees. We
do not see specific threats for the last two categories, but for
the remaining ones it is clear that false positive attacks to shut down
particular industries are within the motives for these attackers, whether for
profit, ideology, or notoriety.

\paragraph{Foreign Intelligence}

This adversary is motivated by allegiance to a nation and unconcerned about the law
in the country where they mount their attack. They may be concerned about their
domestic law but may further receive cover for their actions. They have the
budget that their country deems necessary to mount their attack.

We have already seen state-sponsored disinformation campaigns regarding election
interference~\cite{muller}. Opportunities to selectively suppress or shutdown
particular groups comes readily to the imagination: political
rallies, demonstrations, voters in a particular area, the postal system during
an election, military sites,
and industries supporting the military.

Note that in the context of disinformation, the mere
existence of a technical vulnerability can have disruptive potential in itself,
as an adversary might merely hint at its knowledge of the weakness in order to
erode trust. This effect is accrued for decentralized systems, as it would be
harder to determine whether an attack has even taken place, which can in turn
compound confusion as some would start arguing whether the attacks was
triggered, with little evidence to anchor the discussion.



\paragraph{Cyber Terrorists or Politically-motivated Adversaries}

This adversary is motivated by allegiance to a cause and may or may
not be concerned about the law. Of those willing to mount an attack, they may
further lack a budget. Nevertheless, they may be able to recruit others who share
the same allegiance to a cause to provide the skill required. For example, one
member of an organization may create the popular app that later becomes
weaponized by another member without the first ever intending to weaponize it themselves.

Examples for politically motivated reasons to suppress a particular population
include much from the foreign intelligence category as it relates to election
interference. It also includes motives to support causes, such as disrupting the
operations of slaughterhouses by those against the consumption of meat, and
disruption the operations of oil refineries by those against the production of
greenhouse gas emissions.



\paragraph{Industrial Espionage Agents}

This adversary is motivated by profit and is  concerned about
the law. There would be no national cover or protection for their actions,
though they may exist in less legally stringent nations and conduct their
attacks on either foreign-owned companies  or on industries in foreign nations.
Any large organization will have sufficient budget to mount such an attack.

For those adversaries who face no legal repercussions the benefit of creating a
false outbreak at a rival work site is that they benefit financially by
continuing their own operations, and possibly reputationally depending on the
specific nature of the industry, particularly if an outbreak poses a health risk
to end consumers.


\paragraph{Organized Crime}

This adversary is motivated by profit and unconcerned about the law.
They have sufficient budget to mount the attack so long as the attack itself is
more profitable than the cost. While they cannot write code or
are unwilling to do it without remuneration, they have sufficient budget to have it made.

For these adversaries any effort in a profitable endeavour is sufficient
motivation. Advanced knowledge of any sort of economic disruption is sufficient
to perform criminal insider trading, hoarding then gouging, and extorting
against the threat of an outbreak.

\paragraph{Lesser Criminals and Crackers}

This adversary is motivated by notoriety or curiosity and may or may not be concerned
about the law. For example, they may be a black-hat juvenile insufficiently mature to appreciate the actual
consequences. This adversary is likely to lack a budget, but will have both time
and skill. Were they able to create a popular app they would be able to mount
the attack, and perhaps use it to create a number of fake outbreak hotspots that
prints some message on a map.


%
%
%
%

\section{Future Work}
\label{s:future}

There is future work for this topic. Apps using Bluetooth
need more careful scrutiny and auditing. SDKs that collect Bluetooth information
need particular scrutiny. Our methods identified X-Mode because they did not use
any obfuscation in sending their data, but many SDKs do use various types of
obfuscations. We must examine whenever any app or SDK perform a Bluetooth
scan and examine all the network traffic to scrutinize other SDKs that exhibit the
same behaviour while obfuscating their transmissions.

Additionally, the use of BLE broadcasts can be audited more precisely. This work
simply injected faked observed results, but we did not investigate whether any app or SDK
actually generated their own broadcasts. The broadcasting done over BLE must be
collected and analyzed. In particular, if our fake scan results are
ever later broadcasted we can investigate whether it is attempting to infer MAC
addresses through an echo attack.

Now that proximity tracing apps are more widely deployed, apps with Bluetooth permissions can be
retested to see if their behaviours are changing. Apps that update to now
include \textsc{bluetooth\_admin} when before they did not should be retested as
well to understand the purpose of the change. Furthermore, we did our testing in
Alberta, Canada, which is a jurisdiction that does
not have a decentralized proximity tracing system. Testing of apps in a
jurisdiction may reveal different behaviours, particularly given how the command
and control configuration impacts X-Mode's SDK's behaviour.

\section{Conclusion}
\label{s:conclusions}

In this work, we argued that Bluetooth-based proximity tracing apps are
fundamentally insecure with respect to an attacker leveraging a malevolent app
or SDK. We showed that it is easier than anticipated for an attacker to gain
that capability, and that once there they could launch de-anonymization, false
positive or biosurveillance attacks.  While there are obvious public health
benefits to proximity tracing apps for the purpose of fighting the COVID-19
pandemic which put privacy concerns to the backseat
in this context, our work cautions that the existence of an ecosystem of
surveillance capitalism should not be dismissed, as it is an unfortunate attack
vector that threatens the security and privacy of both centralized and decentralized
proximity tracing systems.

\section*{Acknowledgments}

This work was supported by an NSERC Discovery Grant, a Parex Resources Innovation
Fellowship, the Open Society Foundations, and Luminate. We wish to thank our
anonymous reviewers for their helpful comments.

\raggedright
\bibliographystyle{ieeetr}
\bibliography{references}
\end{document}